\documentclass[fleqn]{article}
\usepackage{espcrc2}

\usepackage{graphicx}
\usepackage[figuresright]{rotating}

\newcommand{\AmS}{{\protect\the\textfont2
  A\kern-.1667em\lower.5ex\hbox{M}\kern-.125emS}}

\title{History of stellar evolution at high redshifts:
Implications for the CMB E-mode polarization}
\author{L.A. Popa \footnote{
also at Institute for Space Sciences Bucharest, R-76900, Romania}
\\
\vspace{0.2cm}
INAF/IASF, Istituto di Astrofisica Spaziale e Fisica
Cosmica Bologna, I-40129, Italy \\
ISS, Institute for Space Sciences Bucharest, R-76900, Romania}

\begin{document}

\begin{abstract}
The epoch of the end of reionization and the Thomson
 optical depth to the cosmic microwave background depend on the power spectrum amplitude on small scales and on the ionizing photon emissivity per unit mass in collapsed halos.
In this paper is investigated the role of the radiative feedback effects for the temporal evolution of the ionizing emissivity.\\
It is shown that the observational constrains
on  hydrogen photo-ionization rate based on
Ly-$\alpha$, Ly-$\beta$ and Ly-$\gamma$
Gunn-Peterson troughs and an electron optical depth consistent
with the latest CMB measurements requires an emissivity of
$\sim$10 ionizing photons per baryon and Hubble time at $z=6$.\\
Through E-mode CMB polarization
power spectrum measurements, is expected that
{\sc Planck} experiment will have the sensitivity
to distinguish
between different histories of stellar evolution.

\end{abstract}
\maketitle

\section{Introduction}
The detailed study of the intergalactic medium (IGM)
is fundamentally important for understanding the
large-scale structure properties and the galaxy formation process.\\
At the epoch of  reionization the collapsed objects began to influence
the diffuse gas in the IGM rendered it transparent to the ultraviolet
photons.
In order to virialize in the potential wells of the dark matter halos,
the gas in the IGM must have a mass
greater than the Jeans mass
$M_J \sim 10^5 M_{\odot}$ at a redshift $z\sim 30$,
corresponding to a virial temperature of $T_{vir} \sim 10^4$K.
Photoionization by the high-redshift UV radiation
background (UVB) heats the low density gas in the IGM before
it falls into the dark matter
wells, strongly reducing the fraction of
neutral hydrogen and helium that
dominate the cooling of the primordial gas at temperatures of $T_{vir}$.

Reionization is an inhomogeneous process that proceeds in a patchy
way. The radiation output associated with the collapsed halos gradually
builds up a cosmic UVB.
At early times,
most of
the gas in the IGM is still neutral
and the cosmological $H_{II}$ regions around the individual
sources do not overlap. At this early stage the gas in the IGM is
opaque to the ionization photons, causing fluctuations
of both ionization fraction and UVB intensity
from region to region.
At the reionization redshift
the $H_{II}$ regions surrounding
the individual sources in the IGM overlap.
During reionization, the number of ionizing photons reaching
 IGM had to be sufficient to ionize every atom
in the universe and to balance the recombination.

From the observational point of view, the study of the
reionization process itself as well as  the properties
of the sources driving it is challenged by a variety of
observational probes.
A powerful  observational probe comes from the Ly$\alpha$
absorbtions spectra of the high redshift quasars
(Becker et al., 2001, Fan et al., 2002, White et al., 2003,
Fan et al., 2004) showing that all
known quasars with $z>6$ have a complete Gunn-Peterson (GP) trough
and a rapid evolving hydrogen neutral fraction compatible with the final
stage of reionization.
An other powerful observational probe is
represented by the high value of the electron scattering optical depth
$\tau$ inferred from
the Cosmic Microwave Background (CMB) anisotropy
measured  by the WMAP experiment
(Kogut et al., 2003; Spergel et al., 2003; Verde et al., 2003)
which requires reionization to begin at $z>14$. \\

In this paper we study the temporal evolution of the ionizing emissivity
in the presence of the radiative transfer effects by using N-body cosmological hydrodinamical simulations at sub-galactic scales.
Detailed discussion of the simulations can be found in
Popa, Burigana, Mandolesi, 2005 (hereafter PBM).
Throughout it is assumed a background cosmology consistent
with the most recent cosmological measurements (Spergel et al., 2003) with
energy density of $\Omega_m=0.27$ in matter, $\Omega_b=0.044$ in baryons,
$\Omega_{\Lambda}=0.73$ in cosmological constant, a Hubble constant of
$H_0$=72 km s$^{-1}$Mpc$^{-1}$, an {\it rms} amplitude of
$\sigma_8=0.84$ for mass density fluctuations
in a sphere of radius 8h$^{-1}$Mpc, adiabatic initial conditions and
a primordial power spectrum with a power-law scalar spectral index
$n_s=1$.

\section{Radiative transfer}

The cumulative UV background flux, $J(\nu_o,z_o)$,
observed at the frequency $\nu_o$
and redshift $z_{o}$,
(in units of 10$^{-21}$ erg cm$^{-2}$ s$^{-1}$ sr$^{-1}$ )
due to photons emitted from redshifts between $z_o$
and an effective emission screen located at $z_{sc}\geq z_o$,
is the solution of the cosmological radiative transfer equation
(Peebles, 1993; Haiman, Rees \& Loeb, 1997):
\begin{equation}
J(\nu_o,z_o)=\frac{c}{4 \pi }\int^{z_{sc}}_{z_o}
e^{-\tau_{eff}(\nu_o,z_o,z)}  \frac{dt}{dz}
 j(\nu_z,z)dz
\end{equation}
where: $j(\nu_z,z)$ is the comoving emission
coefficient
(in units of 10$^{-21}$ erg cm$^{-3}$ s$^{-1}$ sr$^{-1}$)
computed at emission redshift $z$ and photon frequency
$\nu_z=\nu_o(1+z)/(1+z_o)$,
 $\tau_{eff}(\nu_o,z_o,z)$ is
the effective optical depth
at the frequency $\nu_o$
due to the absorption of the residual gas in the IGM
between $z_o$ and $z$ and
$(dt/dz)^{-1}=-H_0(1+z)\sqrt{\Omega_m(1+z)^3+\Omega_{\Lambda}}$
is the line element in the $\Lambda$CDM cosmology.\\
Above the hydrogen ionization threshold
the UV radiation background is processed
due to the absorption of residual gas in the IGM
dominated by neutral hydrogen and helium.
At these frequencies the effective optical depth
is given by (see e.g. Haiman, Rees \& Loeb, 1997):
\begin{eqnarray}
\tau_{eff}(\nu_o,z_o,z)=c\int^z_{z_o}
 \frac{dt}{dz} \kappa(\nu_z,z)dz , \\
\kappa(\nu_z,z)=\sum_i \sigma_i(\nu_z) n_i(z)\, ,\nonumber
\end{eqnarray}
where: $i=(H_{I}\,,He_{I}$) and
$\sigma_i$ and $n_i$
are the cross section and the number
density respectively, corresponding to species $i$.
\begin{figure}

\vspace{-1cm}
\hspace{-1.5cm}
\includegraphics[width=9.cm]{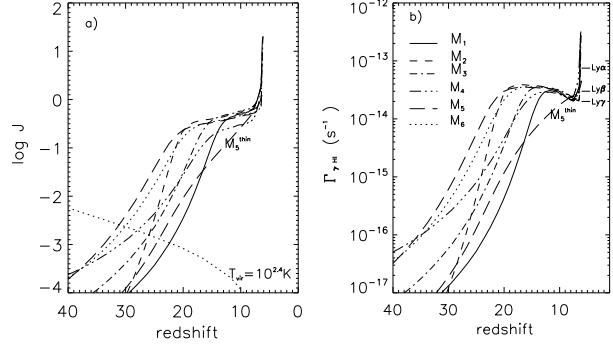}
%\end{flushleft}

\vspace{-1.0cm}
\caption{Left panel: Evolution with redshift
of the UVB flux at $\approx$1 Ryd.
The critical
UVB flux below which ($T_{vir}\leq 10^{2.4}$K)
the star formation is prevented (thick dotted line)
is from Haiman, Abel \& Rees (2000).
Right panel: Redshift evolution of the
hydrogen photo-ionization rate for the UVB models presented
in the right panel.
The observational upper limits  based on
Ly$\alpha$, Ly$\beta$ and Ly$\gamma$ Gunn-Peterson troughs at
$z=6.05$  are from Fan et al. (2002).}
\label{}
\end{figure}

The hydrogen photo-ionization rate, $\Gamma_{\gamma_{H_{I}} }(z)$,
is related  to the UVB flux, $J(\nu,z)$, through:
\begin{eqnarray}
\Gamma_{\gamma_{H{I}}}(z)=\int_{\nu_{H_{I}}}^{\infty}
\frac{J(\nu,z) \sigma_{H_{I}}(\nu)}{h\nu}d\nu\,
%\epsilon_{\gamma_{i}}=\int_{\nu_{th_{i}}}^{\infty}
%\frac{J(\nu,z)\sigma_{i}(\nu)(h\nu-\nu_{th_{i}})} {h\nu} d\nu \,,
%\nonumber
\end{eqnarray}
where $\sigma_{H_{I}}(\nu)$  is the hydrogen photo-ionization (dissociation)
cross-section (Abel et al. 1997)
and $\nu_{H_{I}}$ is the hydrogen
ionizing threshold frequency ($\nu_{H_{I}}\approx$1Ryd).\\
Figure~1 presents (left panel) the evolution with redshift
of the  UVB  flux at $\approx 1$Ryd
obtained for different UVB spectrum  models  (PBM) that constraint the
hydrogen photo-ionization rate  at $z \simeq 6$ to
$\Gamma_{\gamma_{HI}} \simeq  8 \times $10$^{-14}$
s$^{-1}$ (right panel) as indicated by the observational upper limits
based on Ly$\alpha$, Ly$\beta$ and Ly$\gamma$ Gunn-Peterson troughs
(Fan et al. 2002).
In Figure~2 are presented the corresponding
ionization histories with
optical depths in the range $\tau_{es}$=0.05 - 0.1 (PBM).\\
The main effect of the UV radiation spectrum
on the temporal evolution of the ionization
fraction is given by the value of the reionization
redshift and the redshift interval in which the reionization is completed.
As reionization proceeds, the radiative feedback raises
the Jeans mass in the ionized regions,
suppressing the formation of low mass systems.
The net effect is  a decrease of the global
ionization fraction with cosmic
time over a limited period.
\begin{figure}

\vspace{-1.cm}
\hspace{-1.cm}
\includegraphics[width=9.0cm]{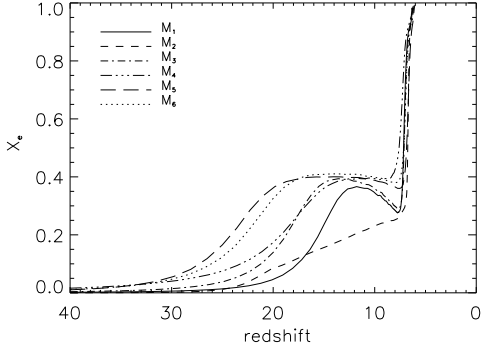}

\vspace{-1cm}
\caption{The evolution with redshift of the ionization fraction
obtained for the UVB models presented in Figure~1 with
electron optical depths $\tau_{es}$=0.05 - 0.1 (PBM).}
\label{}
\end{figure}

\section{Emissivity evolution}

The hydrogen photo-ionization rate  is related
to the ionizing emissivity, $\epsilon_{H_{I}}(\nu,z)$ through:
\begin{equation}
\Gamma_{\gamma_{H_{I}}}(z)=\int^{\infty}_{\nu_{H_{I}}} d\,\nu
\epsilon_{H_{I}}(\nu,z),\lambda(\nu,z)\sigma_{H_{I}}(\nu)\,,
\end{equation}
where $\lambda(\nu,z)$ is the mean free path of the ionizing photons.
Assuming $\sigma_{H_{I}}(\nu)\sim\nu^3$, then $\lambda(\nu,z)$ can
be written as (Miralda-Escud\'e, 2003):
\begin{equation}
\lambda(\nu,z)=\lambda_0\left(\frac{\nu}{\nu_{H_{I}}}\right)^{1.5}\,,
\end{equation}
where  $\lambda^{-1}_0(z)$, the absorbtions probability per unit length
for a photon at $\nu=\nu_{H_{I}}$, is given by:
\begin{eqnarray}
\lambda^{-1}_0(z)=\frac{\sqrt{\pi}}{\lambda_{LL}},
\hspace{0.2cm}
\lambda_{LL}=cH^{-1}(z)
\left(\frac{dN_{LL}}{dz}\right)^{-1}\,,
\end{eqnarray}
with $ H(z)=H_0\sqrt{\Omega_m(1+z)^3+\Omega_v}$.\\
In the above equation $dN_{LL}/dz$ is the Lyman limit system
abundance per unit redshift interval.
Storrie-Lombardi et al. (1994) found
$dN_{LL}/dz=3.3\pm0.6$ at $z$=4 while Madau, Haardt \& Rees (1999)
report a value 1.5 times larger
(see also Miralda-Escud\'e, 2003 and references therein).

The mean ionizing emissivity integrated over frequency,
expressed as the number of ionizing photons
per baryon and Hubble time, can be written as:
\begin{equation}
\begin{small}
\frac{\epsilon(z)}{H(z)n_b}=\frac{\sqrt{\pi}}{c} \frac{dN_{LL}}{dz}
\int_{\nu_{H_{I}}}^{\infty} \frac{J(\nu,z)} {h\nu}
\left(\frac {\nu} {\nu_{H_{I}}}\right)^{-1.5} d\nu,
\end{small}
\end{equation}
where $n_b=2.07 \times 10^{-7}(\Omega_bh^2/0.022)(1+z)^3$cm$^{-3}$ is
the total number density of baryons.\\
Figure~3 presents the evolution with redshift of the mean ionizing
emissivity obtained for  various reionization histories.\\
Figure~4 presents the dependence of the mean ionizing emissivity
on the circular velocity and on the virial mass.
For all  cases,  the experimental constraint on
hydrogen photo-ionization rate
predicts $\sim$10 ionizing photons per baryon and Hubble time at
at $z=6$.
The  electron optical depth of $\tau_{es}=$0.05 - 0.1 requires a mean
ionizing emissivity of earlier low-mass halos
(virial mass of $\sim 10^6M_{\odot}$) of order unity.\\
The emissivity of halos at $z=$6 ($v_{circ}\approx$50 kms$^{-1}$)
is $\sim$35 photons
as obtained by dividing the mean emissivity of 10 photons per
baryon and Hubble time by the
baryon collapse fraction (see Figure~5):
\begin{equation}
F_b(z)=x_e f_{coll}(z,T_h) +
(1-x_e) f_{coll}(z,T_c)\,,
\end{equation}
where $f_{coll}(z,T_h)$ and  $f_{coll}(z, T_c)$ are the collapse fractions
for the regions with the temperature higher than the virial temperatures
$T_h=2.5 \times 10^4$K  and $T_c=10^{2.4}$K (Haiman, Abel \& Rees, 2000;
Barkana \& Loeb, 2001).\\
%__________AICI
In  Figure~5 these numbers are  compared with the number of photons
that can be produced by star formation
assuming that $N_{\gamma}$=2.5$\times$10$^4$ photons per baryon
are emitted by zero-metalicity stars with
$M>20M_{\odot}$
(Onken \& Miralda-Escude, 2004) and  $dN_{LL}/dz=3.6$
(Storrie-Lombardi et al. 1994).
If $f_{*}$ is the fraction of the gas in  halos
that can be converted into stars with $M>20M_{\odot}$ and
$f_{esc}$ is the fraction of the photons that escape into IGM, then
the experimental constraint on
$\Gamma_{\gamma_{HI}} \simeq  8 \times $10$^{-14}$s$^{-1}$
requires $f_{esc}f_{*}\approx 10^{-1.9}$ at $z=6$.\\
The electron optical depth
$\tau_{es}=$0.05 - 0.1 predicts $f_{esc}f_{*}=10^{-2.7}-10^{-1.9}$
for $z=$15 - 6.\\
This values comparable with $f_{esc}f_{*}>$(10$^{-2.4}$,10$^{-2.8}$)
found by Onken \& Miralda-Escud\'e (2004) for the same population
of stars by using an analytical model normalized to
7 photons per baryon and Hubble time at $z=4$ and requiring $\tau_{es}$=0.17
(Miralda-Escud\'e 2003).
\begin{figure}

\vspace{-1.8cm}
\hspace{-1cm}
\includegraphics[width=9cm]{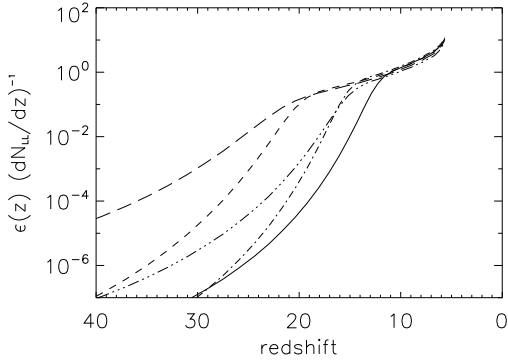}
\vspace{-1.cm}
\caption{Evolution with redshift of the mean ionizing emissivity
(number of ionizing photons per baryon and Hubble time) corresponding to the
reionization histories presented in Figure~2.}
\label{}
\end{figure}
\begin{figure}

\vspace{-2.cm}
\includegraphics[width=9cm]{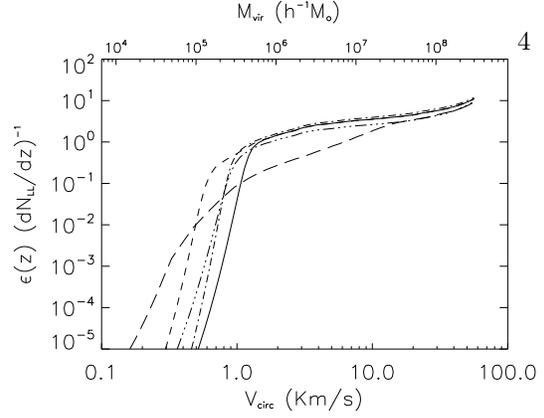}

\vspace{-1cm}
\caption{Dependence on  circular velocity (virial mass) of the mean ionizing emissivity corresponding to the
reionization histories presented in Figure~2.}
\label{}
\end{figure}
\begin{figure}

\vspace{-1.cm}
%\hspace{-1cm}
\includegraphics[width=9.0cm]{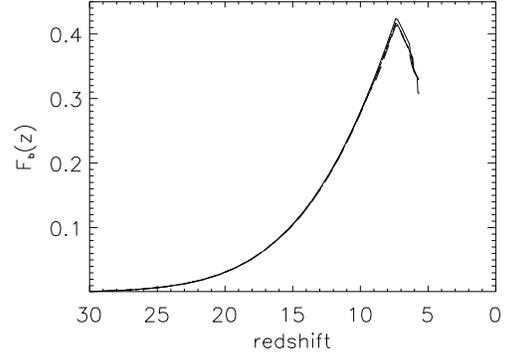}

\vspace{-1.5cm}
\caption{Fraction of mass collapsed into halos
as function of redshift.}
\label{}
\end{figure}
\begin{figure}

\vspace{-0.6cm}
\includegraphics[width=9.0cm]{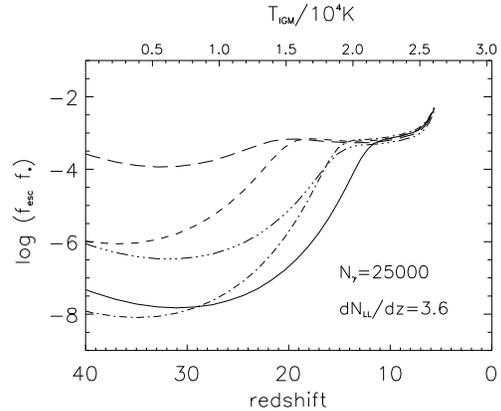}

\vspace{-1cm}
\caption{Evolution with redshift and IGM temperature of the product  $f_{esc}f_*$
corresponding to the
reionization histories presented in Figure~2.}
\label{}
\end{figure}
%__________________________________________________

\section{Implications for the CMB E-mode polarization}

The CMB anisotropy temperature, $C_{T}(l)$, and
E-mode polarization, $C_{E}(l)$,
power spectra corresponding to the
various ionization histories are presented in Figure~6. \\
While $C_{T}(l)$
power spectra are almost degenerated,
the differences in different reionization histories
produce undegenerated signatures on $C_{E}(l)$
power spectra at low multipoles $(l\leq 50)$.
This can be explained by the fact that while polarization
is projecting from the epoch of reionization
at angular frequencies $l=k(\eta_0-\eta_{ri})$
( here $k$ is the wave number, $\eta_0$ and $\eta_{ri}$
are the conformal times today and at the epoch of reionization)
 the temperature is projecting from the (further)
last scattering surface.

It is expected that through E-mode CMB polarization
power spectrum measurements,
the {\sc Planck} experiment will have the sensitivity to distinguish
between different histories of stellar evolution
even they imply the same optical
depth to electron scattering and
degenerated $C_T$ power spectra (PBM).
\begin{figure}

\vspace{-1.cm}
\hspace{-1.cm}
\includegraphics[width=9.0cm]{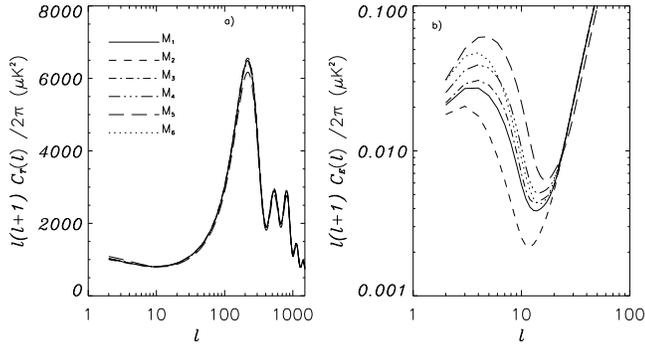}

\vspace{-1cm}
\caption{The CMB angular power spectra, $C_{T}(l)$ and
$C_{E}(l)$ for  the reionization histories
presented in Figure~2.}
\label{}
\end{figure}
%___________________________________________________
%\begin{small}

%\end{small}

\begin{thebibliography}{100}

\bibitem{Abel97}
Abel, T.,  Anninos, P.,  Zhang, Y.,  Norman, M.L., 1997 ,
NewA 2,  181.

\bibitem{Barkana2001}
Barkana, R., Loeb, A. 2001. Physics Reports 349, 125.

\bibitem{Becker2001}
Becker, R.H. et al. 2001. ApJ 122, 2850.

\bibitem{Fan2002}
Fan, X. et al. 2002. AJ, 123, 1247.

\bibitem{Fan2004}
Fan, X. et al. 2004, AJ 128, 515.

\bibitem{Haiman2000}
Haiman, Z., Abel, T., Rees, M., 2000. ApJ 534, 11.

\bibitem{Kogut2003}
Kogut, A. et al. 2003. ApJS 148, 161.

\bibitem{Miralda2003}
Miralda-Escud\'e, J., 2003. ApJ 597, 66.

\bibitem{Madau99}
Madau, P., Haardt, F., \& Rees, M.J. , 1999. ApJ 514, 648.

\bibitem{Naselsky2004}
Naselsky, P., Chiang, L.Y., 2004. MNRAS, 347, 795.

\bibitem{Onken2004}
Onken, C.A., Miralda-Escud\'e, J., 2004. ApJ 610, 1.

\bibitem{popa2005}
Popa, L.A., Burigana, C., Mandolesi, N., 2005. NewA 11, 173 (PBM).

\bibitem{Spergel2003}
Spergel, D.N. et al., 2003. ApJS 148, 175.

\bibitem{lomb1994}
Storrie-Lombardi, L. et al., 1994. ApJ 427, L13.

\bibitem{Verde2003}
Verde, L. et al., 2003. ApJS 148, 195.

\bibitem{White2003}
White, R.L., Becker, R.H., Fan, X., Strauss, M.A., 2003, ApJ 126, 1.

\bibitem{Wyithe2003}
Wyithe, J.S.B., Loeb, A., 2003. ApJ 586, 693.

\bibitem{Wyithe2004}
Wyithe, J.S.B., Loeb, A., 2004. Nature 427, 815.

\end{thebibliography}
\end{document}